\documentclass[
superscriptaddress,
 amsmath,amssymb,
 aps,
twocolumn
]{revtex4}

\usepackage{graphicx}% Include figure files
\usepackage{dcolumn}% Align table columns on decimal point
\usepackage{bm}% bold math
\usepackage[pdftex]{hyperref}
\hypersetup{colorlinks=true,linkcolor=blue,citecolor=blue,urlcolor=blue}
\begin{document}

\preprint{APS/123-QED}

\title{Synthesis of ultrafast waveforms using coherent Raman sidebands}% Force line breaks with \\
%\thanks{A footnote to the article title}%

 \author{Aysan Bahari}
 \email{a.bahari@tamu.edu}
 \affiliation{ Institute for Quantum Science and Engineering, Department of Physics and Astronomy, Texas A \& M University, College Station, TX 77843-4242 USA}
 
\author{Alexandra A. Zhdanova}

\affiliation{ Institute for Quantum Science and Engineering, Department of Physics and Astronomy, Texas A \& M University, College Station, TX 77843-4242 USA}
\affiliation{Exponent, Inc., 149 Commonwealth Drive, Menlo Park, CA 94025}

\author{Mariia Shutova}
\affiliation{ Institute for Quantum Science and Engineering, Department of Physics and Astronomy, Texas A \& M University, College Station, TX 77843-4242 USA}

\author{Alexei V. Sokolov}
 %\email{sokol@tamu.edu}
\affiliation{ Institute for Quantum Science and Engineering, Department of Physics and Astronomy, Texas A \& M University, College Station, TX 77843-4242 USA}

%\collaboration{MUSO Collaboration}%\noaffiliation

%\author{Charlie Author}
 %\homepage{http://www.Second.institution.edu/~Charlie.Author}
%\affiliation{
% Second institution and/or address\\
 %This line break forced% with \\
%}%
%\affiliation{
% Third institution, the second for Charlie Author
%}%
%\author{Delta Author}
%\affiliation{%
% Authors' institution and/or address\\
 %This line break forced with \textbackslash\textbackslash
%}%

%\date{\today}% It is always \today, today,
             %  but any date may be explicitly specified

\begin{abstract}
In this work, we implement a scheme to combine six coherent, spatially separated Raman sidebands generated in single-crystal diamond into a collinear beam. With appropriate phase tuning, this results in a pulse much shorter than the generating pump. We elucidate the characteristics of the synthesized pulse by using an interferometric collinear cross-correlation frequency-resolved optical gating setup (ix-FROG). The beating of the synchronized sidebands results in an additional component in the signal, which we use to optimize the relative phases of our sidebands. In this way, we synthesize and measure visible-range, near-single cycle isolated pulses of approximately 5 fs total pulse duration.
\end{abstract}

%\keywords{Suggested keywords}%Use showkeys class option if keyword
                              %display desired
\maketitle

%\tableofcontents

\section{\label{sec:level1}Introduction}

%The contemporary evolution of ultrashort laser-pulse technology has extended its horizon to a wide variety of areas in physics, chemistry, and biology. This technology is utilized to study the properties of matter and fabricate complex structures for a variety of applications.

% Ultrafast lasers can produce pulses with huge peak powers and power densities which lead to applications such as laser beam machining, multiphoton imaging, and the production of high-frequency electromagnetic radiation (such as x-rays)\cite{Gavin2016}. 

The need to understand and control electron motion on faster and faster time scales \cite{Zewail2000,Corkum2007b} has driven ultrashort laser pulse technology towards shorter and shorter pulses. Ultrashort laser pulses are often generated with mode-locked lasers by passive and active mode-locking followed by various compressive techniques \cite{Weiner2009,Nisoli97,Chan2011}. One of the most popular methods of generating attosecond pulses is high harmonic generation (HHG)  \cite{Christov1997,Farkas1992}. However, there are several intrinsic limitations to this technique, including its fundamental inefficiency, small energy throughput and the difficulty of controlling and maintaining single-cycle x-ray pulses \cite{Harris1993a}. Another popular method utilizes fiber-generated supercontinua split into multiple branches, compressed, and recombined to generate very short, single or sub-femtosecond pulses \cite{Krauss2010,Hassan2016,Wirth2011}. While this approach has been shown to generate relatively high power (several hundred $\mu$J) and ultrafast (on the order of a single cycle or less) pulses, the pulse power is fundamentally limited by the power output of the generating fiber. Another approach utilizes noncollinear optical parametric chirped pulse amplifiers to produce pulses of several hundred mJ with pulse durations down to the approximately 6-7 fs range \cite{Herrmann2009}. The results reported in this work are, in principle, only limited in power by what the pump laser can produce and in bandwidth by the total generated bandwidth of the Raman process (which routinely spans 350-1100 nm). Our work is based on a technique dubbed ``molecular modulation".

In the past few decades, the molecular modulation technique has played an essential role in producing such short pulses (femtosecond and attosecond) in the optical region \cite{Wang2010,Yavuz2007, Gold2014}. This technique is based on the frequency modulation of a laser pulse propagating through a coherently vibrating ensemble of molecules, which results in the generation of multi-color sidebands that are spaced by the vibration frequency of the molecule. These sidebands, dubbed ``Raman sidebands", are all produced coherently, in a phased manner, potentially generating sub-femtosecond pulses with any desired pulse shape where the electric field is not limited to a quasi-sinusoidal oscillation\cite{Sokolov2005}. Using this technique, Sokolov \textit{et al.} demonstrated the synthesis of a pulse train of nearly single-cycle waveforms in the adiabatic excitation of the D$_2$ molecule \cite{Harris2003}. In a similar direction, Zhavoronkov and Korn generated pulses with duration below 4 fs \cite{Zhavoronkov2002} by utilizing  a hollow-core waveguide filled with an impulsively pre-excited Raman-active gas, while Suzuki \textit{et al.} generated an octave-spanning Raman comb from single-frequency lasers in gaseous parahydrogen\cite{Suzuki2008}.

The molecular modulation technique is not limited to gaseous media and has been extended to nonlinear solids such as diamond and PbWO$_4$ crystals\cite{Zhi2008,Zhdanova2015b}. In solid-state media, it is possible to use femtosecond pulses (instead of picosecond or nanosecond) to produce coherent Raman sidebands, opening up the possibility for synthesizing single-cycle, isolated, visible pulses. Similar to gaseous media, collinear interaction is possible in solids, but due to the dispersion of the medium, result in a situation where sideband generation is optimized at a certain non-zero crossing angle. This configuration is achieved by crossing the pump and Stokes input beams at a specific angle inside the crystal \cite{Zhi2008}. As a result, the generated sidebands are produced at different output angles, and therefore, additional techniques are required to recombine these sidebands and synthesize ultrashort pulses \cite{Wang2014,Wang2015a,Zhi2013}. In this work, we apply a novel scheme to generate and characterize a few-cycle pulse in a single polychromatic beam. Our setup uses dichroic mirrors to recombine the spatially separated sidebands, and the resulting waveform is characterized via a technique we dub interferometric cross-correlation frequency resolved optical gating (ix-FROG). This technique is a combination of cross correlation FROG (XFROG) \cite{Trebino2000} and interferometric FROG \cite{Stibenz2005, Amat-Roldan2004,Zhi2011, Hyyti2017}. 

\newpage
\onecolumngrid
\begin{figure*}[t]
	\begin{center}
		\includegraphics[scale=0.6]{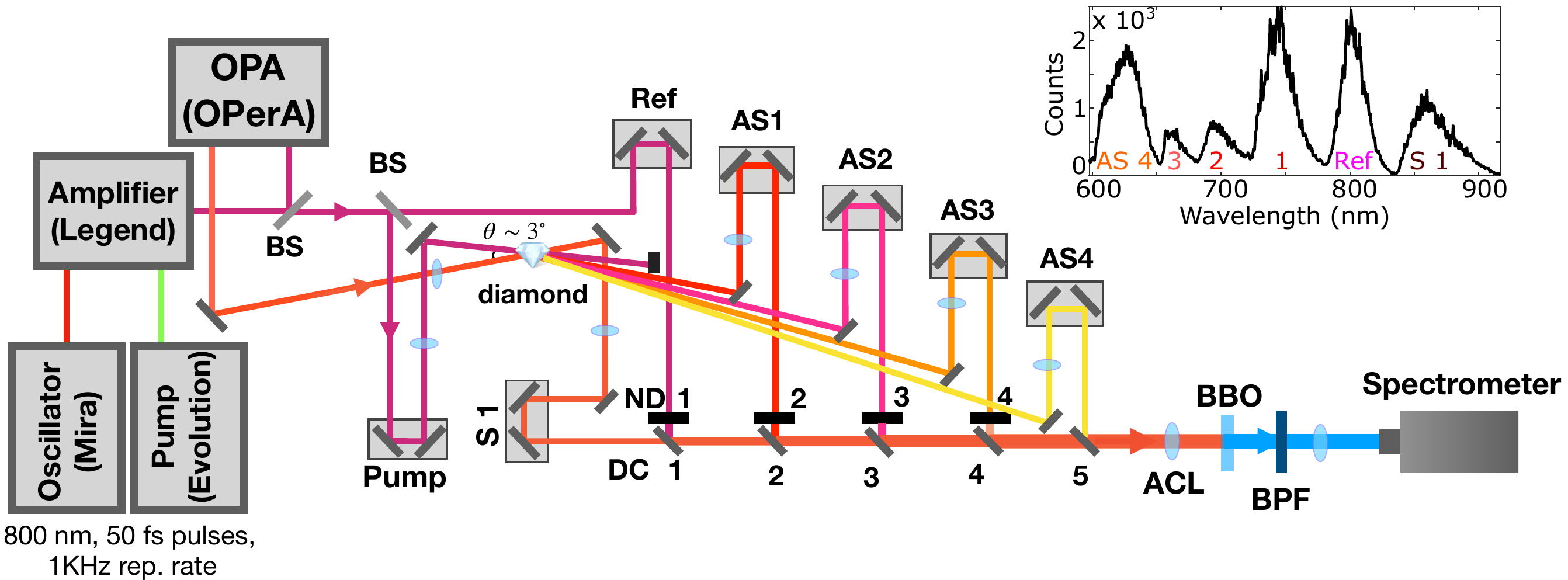}
		\caption{Our experimental setup to synthesize a sub-5-fs FWHM pulse. The black bars (ND 1-4) before each DC mirror represent variable neutral density filters which we use to adjust the intensity of each beam to match the intensity of AS4. Abbreviations for the optical elements: BS, 50/50 beamsplitter, DC 1-5, dichroic mirror 1-5, ACL, achromatic lens,  BBO, beta-barium-borate crystal, BPF, band-pass filter. Inset: recorded average spectrum of our synthesized pulse, spanning from AS4 to S1 ($\sim300$ nm of bandwidth). The bands are not optimized for phase recombination.}
		\label{fig:synthesissetup}
	\end{center}
\end{figure*}
\twocolumngrid

The cross-correlation part of the technique allows us to retrieve the waveform of each sideband, while the interferometry allows us to determine the phase of each beam with respect to the others. Although several experimental factors limit our current synthesized pulse energy and duration, a number of adjustments can be applied to the setup to overcome some of these limitations. For instance, increasing the pump energy without reaching the damage threshold of the Raman crystal can significantly increase the energy of the sidebands, and therefore, the energy of the synthesized ultrashort pulses. Focused on the niche of molecular modulation technique, our Raman source presents a broad range of applications, for example, in optical coherent tomography, ultrafast spectroscopy, and precision metrology. Moreover, our technique can be potentially used in generating coherent radiations in the UV and IR spectral regions, where laser sources are not readily available. This work is a step toward expanding the flexibility and applicability of our technique, while paving the way towards ever-shorter pulses.  

\section{Experimental setup}
\label{sec:setup}
Fig. \ref{fig:synthesissetup} shows the experimental setup we used to synthesize and characterize our ultrashort pulse. We split the main laser line at $800$ nm with a low-group-delay-dispersion (GDD) beamsplitter. We use one leg (called pump in CARS terminology) in conjunction with the second harmonic of the idler out of the OPA (Coherent OPerA) at $870$ nm (called Stokes, S1 in CARS terminology) to stimulate the $1332$ cm$^{-1}$ Raman line of a $0.5$ mm thick, single-crystal diamond. The OPA is pumped directly from the main laser line and is seeded by white light generated from a fraction of pump in a sapphire plate; hence, S1 has a carrier-envelope-phase which is, in principle, related to the carrier-envelope-phase of pump.

We combine the two beams at a $\sim 3^{\circ}$ angle, focusing each individually with a $50$ cm lens (S1) and a $30$ cm lens (pump). The S1 beam profile is optimized with an iris prior to focusing. The average pulse energy in each beam is $18.28$ $\mu$J (pump) and $1.86$ $\mu$J (Stokes) as measured by a Coherent PM10 power meter. This configuration produces many orders of Anti-Stokes (AS) Raman sidebands. These sidebands are essentially frequency-shifted copies of the original femtosecond pulses. After exiting the crystal, we collimate each sideband individually. Wherever possible, thin lenses were used to avoid adding substantial dispersion. However, adding the dispersion of the lenses has a minimal effect on the final synthesized pulse so long as the relative phase between each band is properly adjusted as part of the interferometric setup. This is because, despite the dispersion of the lenses, the phase in the most intense part of each beam remains relatively flat (smooth, $<2\pi$ change across $100$ fs), as shown in Fig. \ref{fig:xfrogs}. Moreover, the synthesized pulse duration is mainly affected by the total frequency span of the sidebands which participate in the synthesis, and this factor is not affected by the dispersion added from the lenses, ND filters, etc. used in our setup.

After collimation, each band is aligned to a separate delay line to allow for full control of phase and flexibility of position. These bands are then recombined with the remains of S1 after the diamond and the other leg of the split main laser line (dubbed ``Reference''). We used commercially available dichroic mirrors for this recombination, and the recombined spectrum is shown in the inset of Fig. \ref{fig:synthesissetup}. 
Some sidebands' spectra have reduced bandwidth due to the cutoff frequencies of the dichroic mirrors, as is consistent with the retrieved pulse shapes in Fig. \ref{fig:xfrogs}. Full information on the collimation lenses, dichroic mirrors, and translation stages used in the setup are available in Table \ref{tab:synthesisparts}. We found the power of the least powerful sideband (i.e. AS4) after recombination to be $6.5$ nJ (in comparison to AS1's $200$ nJ of power); the power of all other sidebands was reduced with ND filters to match this power.

After dichroic recombination, we used a single achromatic doublet lens to focus the beams into a $10$ micron beta-barium-borate (BBO) crystal to characterize the resultant pulse. Specifically, second harmonic and sum frequency signals of the sidebands are generated in the BBO in the spectral range $340 - 450$ nm,  with intensity dependent on phase relation between individual sidebands in an interferometric configuration. After exiting the crystal, the fundamental sideband frequencies are filtered depending on which spectral region is under investigation. For signals above $390$ nm, a lone Thorlabs FGB25 UV band-pass filter (BPF) was used, otherwise, an additional Thorlabs FGUV11 BPF was added. 

\newpage
\onecolumngrid
\begin{table*}
\caption{\label{tab:synthesisparts}Part numbers for the dichroic mirrors and translation stages; focal lengths for the collimation lenses used in this setup. Part numbers which start with ``TL'' correspond to Thorlabs part numbers, ``NP''  - Newport, ``EO'' - Edmund Optics, and ``SR'' - Semrock.}
\begin{ruledtabular}
\begin{tabular}{ccccc}
 %&\multicolumn{2}{c}{$D_{4h}^1$}&\multicolumn{2}{c}{$D_{4h}^5$}\\
 Band&Delay stage&Collimating lens(es)& DC mirror\\ \hline
			S1 & NP 423 series & $50$ cm & N/A \\ 
			Ref. & TL LNR25ZFS & N/A & EO 69-895 \\
			AS1 & NP GTS150 & $40$ cm & SR FF776-Di01 \\
			AS2 & NP 443 series/TL PAS005 & $15$ cm \& $-10$ cm & SR FF735-Di02 \\
			AS3 & TL LNR25ZFS & $25$ cm \& $-10$ cm & SR FF685-Di02 \\ 
			AS4 & NP 423 series & $30$ cm \& $-10$ cm &SR  Di03-R635-t1 \\ \hline
\end{tabular}
\end{ruledtabular}
\end{table*}
\twocolumngrid

The resulting UV signal was focused with a $7.5$ cm lens into a multi-mode fiber and analyzed via a spectrometer (Ocean Optics HR4000). Different nonlinear signals result from blocking or unblocking different bands, as discussed in the following sections. 

It is important to note that our setup currently has no active stabilization or noise jitter suppression. While this does preclude the use of the setup presented herein from single shot measurements, we use the results presented in Fig. \ref{fig:resultsbeams} to show that the setup is stable enough to repeatedly and reliably measure intensity fluctuations on the sub-single-femtosecond scale. Hence, our setup allows us to take repeated interferometric measurements which we can then average to reduce the noise. Active suppression of the noise would result in more consistent results, and can be implemented in future iterations of this setup.

\section{ix-FROG pulse retrieval}
\begin{figure*}[t]
	\begin{center}
		
		\includegraphics[width=0.8\textwidth]{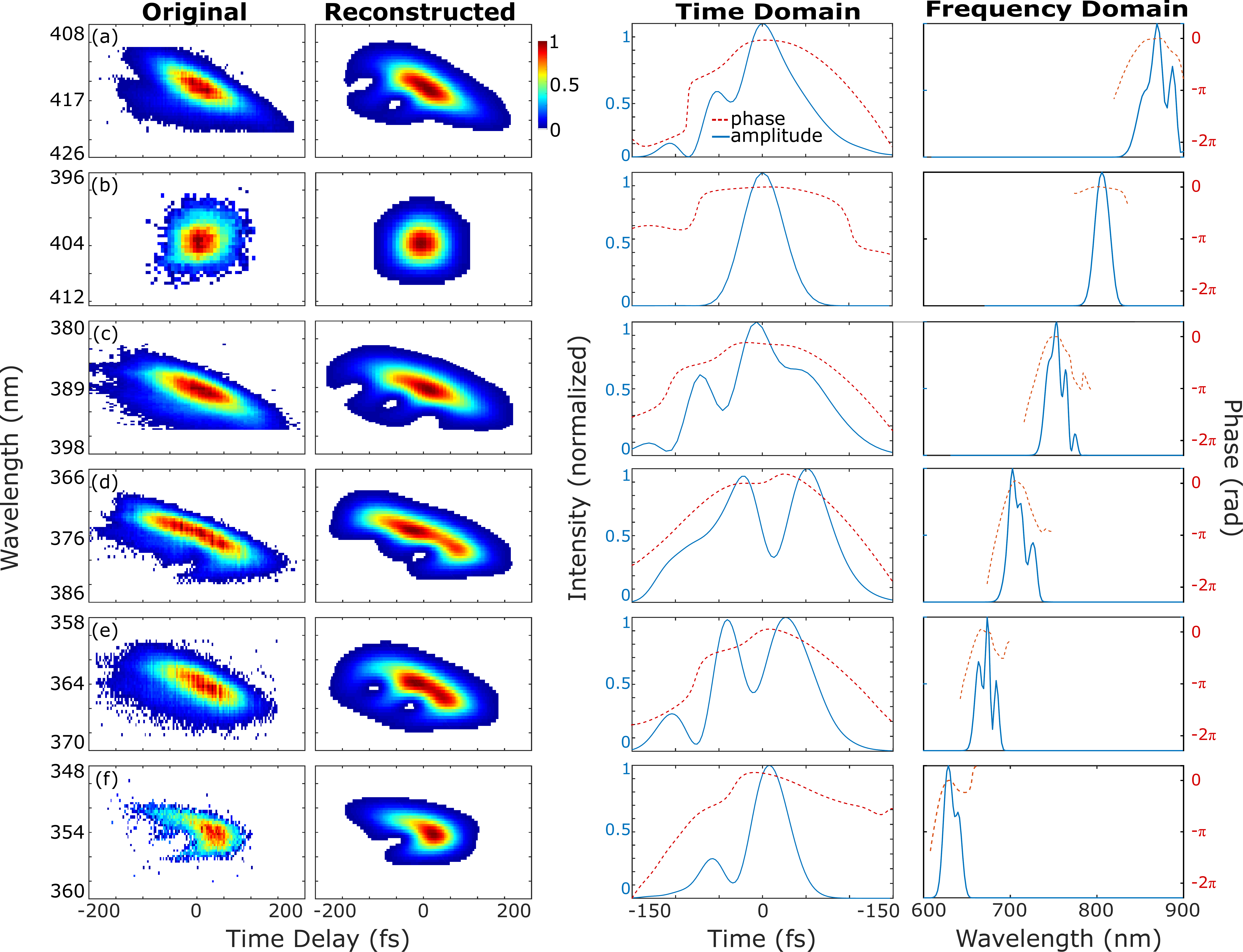}
		\caption{Experimental XFROG spectrograms of all beams employed in our setup, using Reference as the known pulse to gate (a) S1, (b) itself in an SHG FROG configuration, (c)-(f), AS1-4. The columns show the measured and reconstructed spectrograms, as well as a plot of the retrieved temporal pulse shapes (phase and amplitude).}
		\label{fig:xfrogs}
	\end{center}
\end{figure*}

The first step in our ix-FROG technique is to record standard XFROG spectrograms for each beam. These spectrograms were taken and recorded individually (i.e. by blocking and unblocking various beams) to ensure no extra noise or background from the interference terms described below. However, it is also possible to take all spectrograms simultaneously by removing the resultant SFG background and filtering out the higher order interference terms, as is done in \cite{Amat-Roldan2004}. We used the 800 nm band dubbed Reference as our known pulse; we first characterized this pulse with a separate homebuilt SHG-FROG setup. This setup utilized a very small ($<3^{\circ}$) recombination angle and $10$ micron BBO, achieving results consistent with what we expect from our commercial laser amplifier. This SHG-FROG trace is shown in Fig. \ref{fig:xfrogs}(b).

Collinear XFROGs were then taken with all dichroics and filters in place; we varied the delay of Reference with respect to all other beams as our gating pulse. We used the standard XFROG algorithm provided on Dr. Trebino's website for pulse retrieval \cite{Trebino2016a}. Fig. \ref{fig:xfrogs} displays the results. We successfully retrieved the Raman sideband pulse shapes with $< 2\%$ RMS difference between the experimentally obtained spectrograms and FROG-reconstructed spectrograms, indicating very good retrieval. Severe distortions in the pulse shapes of AS2 and AS3 stem from the variable group-delay-dispersion (GDD) in their respective dichroic mirrors close to the cut-off frequency \cite{Semrock2017}.

\twocolumngrid

\begin{figure*}[t]
	\begin{center}
		
		\includegraphics[width=0.8\textwidth]{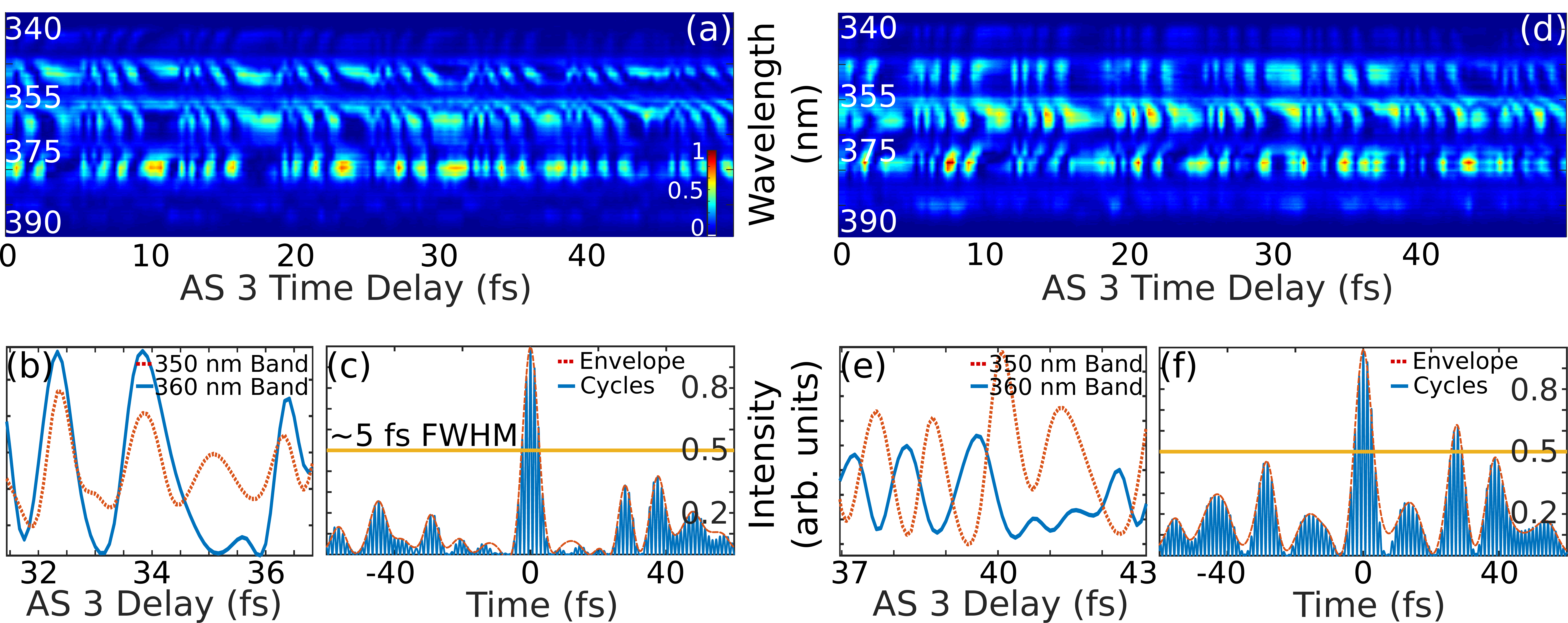}
		\caption{
       	Comparison of experimentally measured ix-FROG traces for the waveforms formed by AS2 being in-phase (a--c) and out-of-phase (d--f) with the other sidebands.
          (a,d) Full spectrograms (spectrum of the sidebands as a function of AS3 sideband delay). 
           (b,e) Cuts of the spectrogram at the 350 nm band (red dashed line) and at the 360 nm band (blue dashed line). When AS2 is in phase with the rest of the sidebands (in-phase waveform, (b)), the maxima of the interference fringes in the 350 nm and the 360 nm bands of the ix-FROG trace are aligned. When AS2 is out of phase with the rest of the sidebands (out-of-phase waveform, (e)), the maxima of the interference fringes in the 350 nm band correspond to the minima in the 360 nm band. 
          (c,f) Reconstructed temporal profile of the in-phase (c) and out-of-phase (f) waveforms; assuming a carrier-envelope-phase of $0$.
       }
		\label{fig:resultsbeams}
	\end{center}
\end{figure*}

Once collinear XFROGs were individually taken, a full spectrogram was obtained by unblocking all beams. This full spectrogram shows clear interference on all bands, at a period roughly equal to $1.2$ fs or $1/f$, where $f$ is the frequency of the band under question. This is expected by the interaction and coherence between all nonlinear signals. For example, in Fig. \ref{fig:resultsbeams}(a, d), the $350$ nm band represents the interference between the second harmonic of AS3 with the sum-frequency of AS2 and AS4. Similarly, the $360$ nm band represents the interference between the second harmonic of AS2 with the sum-frequency of AS1 and AS3, as well as the interference between Reference and AS4. Detailed descriptions of the terms which contribute to the interference are given in \cite{Zhi2011}. 

In this setup, we can vary the delay of any of the sidebands to obtain such an interferometric picture. However, as proof-of-principle, we only examine the results of varying the delay of AS3, as shown in Fig. \ref{fig:resultsbeams}(a,d). To make the details of the interference patterns more visible, when plotting the spectrograms we subtracted the constant background and interpolated to a 0.17 fs step size (1/4 of the actual step size) using standard spline interpolation. By adding or removing phase from a particular beam, the interference channels shift with respect to each other, as discussed further in Section \ref{sec:results}.  In essence, the ix-FROG technique measures relative phase between the sidebands. Combining it with the XFROG pulse measurement technique, we find the shape of the multi-sideband waveform. 

Note that in our proof-of-concept setup we do not use carrier-envelope phase (CEP) stabilization of pump or Stokes pulses. Since the phase of each anti-Stokes sideband depends on phases of both pump and Stokes pulses, it is reasonable to ask if the shape of our synthesized waveform is stable from one shot to another. To see why that is the case, let us denote the CEP of the pump pulse as $\phi_p$ and CEP of the Stokes pulse as \cite{Sokolov2005,Chen2008}:
\begin{equation} \label{eq_phin}
\phi_n = \phi_p + n \left( \phi_p - \phi_S \right)
\end{equation}
while the frequency of $n$-th order sideband is given by
\begin{equation} \label{eq_freq_n}
\omega_n = \omega_p + n \left( \omega_p - \omega_S \right)
\end{equation}
(where $\omega_p$ is the frequency of pump and $\omega_S$ is the frequency of Stokes). Mathematically, the frequency and phase of pump and Stokes pulses also follow Eqs. \eqref{eq_phin} and \eqref{eq_freq_n} with $n=0$ and $n=-1$ respectively, so in the experiment we combine sidebands with indices $n=-1,0,1,\dots,4$. Equations \eqref{eq_phin} and \eqref{eq_freq_n} show that CEPs of all combined bands are linear in frequency $\omega_n$: 
\begin{equation}
\phi\left(\omega_n\right) = \phi_p + \frac{\omega_n - \omega_p}{\omega_p - \omega_s} \left(\phi_p - \phi_S \right)
\end{equation}

It follows then from Fourier theory that random changes in $\phi_p$ and $\phi_S$ affect the synthesized waveform in two ways: 1) change in CEP of the waveform and 2) random shift in time. In other words, even though both $\phi_p$ and $\phi_S$ are random, the shape of the envelope of the synthesized pulse does not change (although its precise arrival time does). Essentially, the main knob affecting the pulse shape in our waveform synthesis setup is a constant, additional phase between the sidebands, which depends on $n$, but not on CEP $\phi_p$ or $\phi_S$. This extra phase between sidebands is controlled via adjustment of the optical paths of the bands. Finally, as discussed in Section \ref{sec:setup}, our Stokes beam is generated from the pump beam. Hence, in principle, since the carrier-envelope-phase of our setup is only affected from shot to shot by the CEP of our pump laser, our setup is compatible with standard 

\newpage
\onecolumngrid

\twocolumngrid
\noindent methods of CEP stabilization \cite{Telle1999}.  Note that our technique is decoupled from the standard CEP stabilizing methods such that CEP stabilization does not affect the results of this manuscript.

\section{Results}
\label{sec:results}

Once all pulses are overlapped in space and time, a waveform is synthesized throughout the beam by the coherent addition of the individual sidebands. As discussed previously, the shape of the synthesized waveform is controlled by the phase relationships between sidebands. This is seen qualitatively in the ix-FROG traces of the waveforms which examine the phase of AS2 with respect to the other sidebands  (Figs. \ref{fig:resultsbeams}(a) and \ref{fig:resultsbeams}(d)). For the in-phase waveform the interference fringes in various bands of the ix-FROG spectrogram are aligned, i.e. have maxima at the same AS3 delay (Fig. \ref{fig:resultsbeams}(b)). In Fig. \ref{fig:resultsbeams}(c)-(f), we add an extra phase to AS2 by moving its piezoelectric stage slightly forward.

\noindent This results in an out-of-phase waveform where the spectrogram interference fringes are anti-aligned -- maxima at 350 nm and 370 nm bands correspond to minima at 360 nm band and vice versa (Fig. \ref{fig:resultsbeams}(e)).

By putting all pulses in phase (i.e. by stopping on a bright spot in Fig. \ref{fig:resultsbeams}(a)) we obtain an isolated 5 fs pulse, as shown in Fig. \ref{fig:resultsbeams}(c). Setting all beams in phase is essential for optimal synthesis to take place; if AS2 is out of phase with all the other beams, the temporal contrast of the main pulse with respect to the pre- and post-pulse worsens; technically, a 30 fs FWHM pulse is obtained (even though the FWHM of the main pulses increases only slightly to 6 fs). This is also shown in Figs. \ref{fig:resultsbeams}(c,f).

A detailed inspection of the ix-FROG traces shows more structure than simply in and out of phase interference fringes. For instance, periodically, fringe visibility drops dramatically for about $\sim 4$ fs of the AS3 delay. Fringes in the 360 nm band are also tilted, i.e. fringe maxima at different wavelengths within a band correspond to different AS3 time delays. Our qualitative model of ix-FROG does not account for any of these effects, and these will be the subject of a future publication. 

\section{Conclusions}
We have demonstrated that, by using our ix-FROG technique, we can synthesize and measure an isolated (3:1 signal-to-noise) 5-fs pulse. This setup is only limited in bandwidth and power by what is produced in the Raman process and can be scaled in both to produce single-cycle isolated pulses at a much higher power, ideal for studying ionization and other processes on the single-femtosecond time scale.

\vspace{-0.6cm}
\acknowledgments
 We gratefully acknowledge Rick Trebino, his group, and Peter Zhokhov for FROG and simulation related discussions. This work is supported by the National Science Foundation (NSF) (CHE-1609608); the Robert A. Welch Foundation (A1547); and the Office of Naval Research  (Award N00014-16-1-2578). A.B. and A.A.Z. are supported by the Herman F. Heep and Minnie Belle Heep Texas A\&M University Endowed Fund held/administered by the Texas A\&M Foundation.
 
\bibliographystyle{unsrt}
\bibliography{Ref}% Produces the bibliography via BibTeX.

\end{document}